# THE INFLUENCE OF THE METEOROLOGICAL FORCING ON THE RECONSTRUCTIONS OF HISTORICAL STORMS IN THE BLACK SEA

**Vasko Galabov**, **Anna Kortcheva**

National Institute of Meteorology and Hydrology- Bulgarian Academy of Sciences, **Bulgaria**
Corresponding author: Vasko.galabov@meteo.bg

**ABSTRACT**

The present article is a study of the applicability of different sources of meteorological forcing for the coastal wave and storm surge models, which provide the operational marine forecasts for the coastal early warning systems (EWS) and are used for reconstructions of historical storms. The reconstructions of historical storms are one of the approaches to the natural coastal hazard assessment. We evaluate the importance of the input meteorological information for the mentioned types of coastal models. For two well documented historical storms, that caused significant damages along the Bulgarian coast we simulate the significant wave heights and sea level change, using SWAN wave model and a storm surge model. The wind and mean sea level pressure fields, which are used in the present study, are extracted from the ERA Interim reanalysis of the European Center for Medium range Forecasts (ECMWF) and from the output of the high resolution limited area numerical weather prediction model ALADIN. The overall conclusion is that for the successful historical storms reconstructions ERA Interim and ERA40 reanalysis are valuable source of meteorological forcing, but due to their limitations in terms of spatial and temporal resolution, it is recommended to produce a higher spatial and temporal resolution meteorological fields, using dynamical downscaling of the reanalyzed data.

**INTRODUCTION**

Natural hazards continue to be one of the major causes for the loss of human lives and economic damages, damages to the ecosystems and infrastructure (with the associated decrease in the quality of life of the population in the areas affected by those hazards). It is important to mention here, that the vulnerability to natural hazards is a major issue not only in the developing world, but also in the countries with the highest economic development. Recognising such threats and also the possibility, that the climate changes could possibly cause intensification of some dangerous natural events, the European Union (EU) continues to increase the support to scientific studies and other types of projects, leading to an increasing resilience of the society to the natural hazards. The present study is a part of the on-going EU FP7 project IncREO (Increasing Resilience through Earth Observations). Partners in this project are: Spotimage (Astrium GEO)-

France, Geomer GmbH Germany, Geo Ville information systems- Austria, Infoterra GmbH- Germany, University of Twente- Netherlands, Romanian Space Agency, UNESCO, the French national weather service Meteo- France and the Bulgarian National Institute of Meteorology and Hydrology (NIMH- BAS). The goal of the project is to provide to the authorities, responsible for the disaster management, civil protection and risk prevention, solutions that will help to improve the preparedness and planning for highly vulnerable areas and noticeable effect of the climate changes. The project accents on the multirisk approach. Cases of areas, vulnerable to dam failures, coastal hazards- storm surge and high waves, floods, earthquakes, landslides are selected. The role of Meteo- France and NIMH- BAS in this project is to study the coastal hazards vulnerability due to strong winds, high waves and storm surges. The use cases are the coasts of France and Bulgaria, but taking into account that by this choice we study the coastal vulnerability in the Mediterranean coast of France, Atlantic coast of France and the Bulgarian coast, the results are highly transferable.

For the purposes of coastal hazards assessments there are different approaches, some of which are focused on long term continuous wave and storm surge hindcasts the on-going project IncREO studies selections of historical severe coastal storms during the last decades by the use of numerical modelling. The use of numerical modelling is especially important in the case of the Black Sea coast due to the lack of instrumental measurements of waves (except some episodic campaigns and experiments). One of the most important issues of such studies is the choice of meteorological input to such numerical models. In this article we present simulations of some well documented historical storms in the Black Sea coast and evaluate the usability for wave and storm surge hindcast of one of the most widely used sources of meteorological fields such as ERA Interim reanalysis [1], provided by ECMWF.

**MATERIALS AND METHODS**

The numerical models, which we use in this study, are a nearshore numerical wave model and a storm surge model, but it is important to mention, that the conclusions are not dependant on the models choice, due to fact, that the meteorological input is crucial for all such models accuracy. The wave model that we use is the SWAN wave model, developed by TU-Delft, Netherlands [2]. SWAN has been implemented and validated by several research groups in the Black Sea: E. Rusu implemented SWAN and studied the wave current interactions [3], wave energy for selected cases [4], and the wind generation and dissipation parameterisations [5]; Akpinar et al [6],[7] validated SWAN and estimated the wave energy potential of the Black Sea; L. Rusu also implemented SWAN in the Black Sea[8], and compared different third generation models [9]; Fomin et al tested SWAN in the shallow bays of Crimean peninsula [10]. Details of the Bulgarian operational implementation of SWAN can be found in [11] and [12].

The storm surge model, that we use, is a development of Meteo France [14], adopted and customised for the Black Sea by Mungov and Daniel P [13]. The model is a two dimensional and takes into account the advection, horizontal turbulence, nonlinear bottom and surface friction. Mungov and Daniel conclude in [13] that the model gives accurate predictions of the sea level with some underestimation, for which they suggest, that it is due to meteorological input fields. The model integrates the currents taking the mixed layer depth as liquid bottom.

The meteorological inputs (10m wind fields and mean sea level pressure fields), that we evaluate are extracted from ERA Interim reanalysis of ECMWF [1]. Their spatial resolution is 0.75° and the temporal resolution- 6h. Using ERA Interim we simulate the storm of February 1979 (the most severe recorded storm at the Bulgarian coast) and also the storm of 06-08 February 2012 (see [11] for more details) and also a storm (February 2003) with a short but extreme increase of the significant wave height in deep waters for the Eastern Black Sea coast in order to compare the models with instrumental observations from the buoy nearby Gelendzhik, Russia (data are provided by the Russian oceanographic institute Shirshov- available online at www.coastdyn.ru). For the last two cases we compare the simulations, using ERA Interim with simulations by the use of meteorological input from the Bulgarian implementation of the high resolution regional atmospheric model ALADIN (for the details about the Bulgarian implementation of ALADIN model see [15]). The available model output for 2003 is with spatial resolution of 0.25° and temporal- 6h, while for the case of 2012 the spatial resolution is 0.125° and the temporal- 3h. For the case of 1979 we simulate the waves and the storm surge (due to the documented extreme surge) and for the other two cases only the waves.

Fig.1 presents the locations of the sites of the Bulgarian coast, which are mentioned in text of the present article.

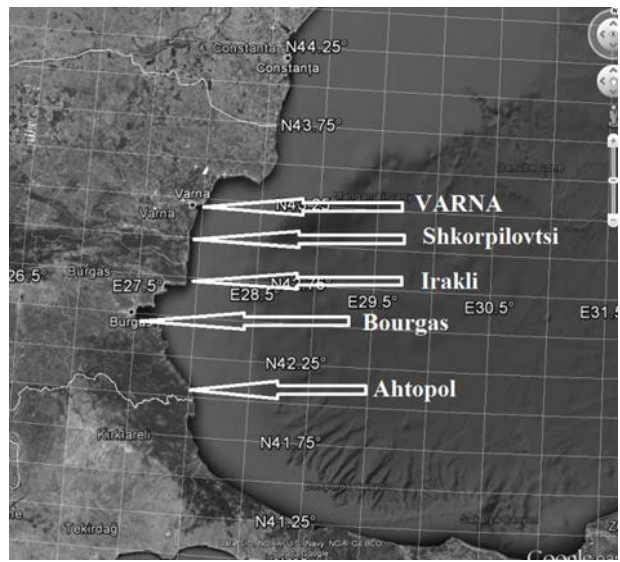

**Fig.1.** Locations of some places of interest, which are mentioned in the study- towns: Varna, Bourgas and Ahtopol, the village Shkorpilotsi (measurements site of the Institute of Oceanology of BAS), Irakli beach- site of a mareograph station.

**THE CASE 1: THE STORM OF FEBRUARY 1979 (THE BULGARIAN STORM OF THE CENTURY)**

The first reconstructed storm, using ERA Interim atmospheric input, is the storm of February 1979. This storm is the most frequently studied storm, that affected the Bulgarian coast according to the research of literary sources by Trifonova et al [16]. Belberov et al presented a reconstruction of the storm [17] by numerical models. This storm is considered due to the fact, that it caused the most severe damages to the Bulgarian coast of all recorded storms and it is considered as a "storm of the century" (100 years event). During the storm the sea level in the tide gauge station at Irakli beach reached 1.43 above the mean (according to [16] - the data for the sea levels are according to that study). The data from the measurements in Varna and Burgas are not available, due to the fact, that the mareograph equipment wasn't designed to measure sea level values more than 1.5m and they has been flooded and damaged. The significant wave height near the village of Shkorpilovtsi (see fig.1) reached at least 5.8m at 15m depth and it is not clear if that is the maximum, due to the fact, that the storm destroyed the wave measurement equipment and the pier with the other measurement equipment of the Bulgarian Institute of oceanology (IO-BAS).

We simulate the storm, using SWAN wave model and the storm surge model mentioned in the previous chapter, forced by ERA data, considering also the breaking waves contribution (wave setup) to total sea level rise. Fig.2 shows the simulation of the significant wave height for the position of the pier of IO-BAS at 15m depth.

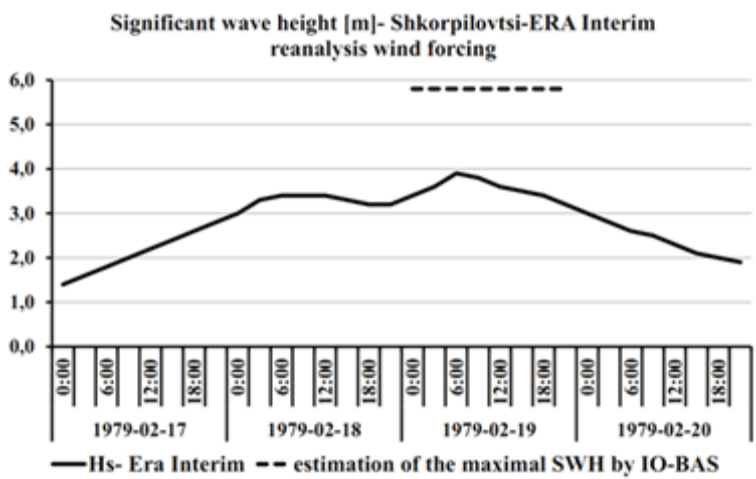

**Fig.2.** SWAN model simulation of the significant wave height near the village of Shkorpilovtsi during the storm of February 1979.

The simulated significant wave height is with a maximal value of 4m, while the actual was at least 5.8m and therefore the underestimation is 50%. Our explanation is that the reason for this is the coarse spatial resolution of the reanalysis and the lack of enough remote sensed data, assimilated in the models during the 70's. A serious problem is that in order to estimate the impact of such extreme waves, the significant wave height is not the only important parameter. The wave energy flux is also very important in order to estimate the impact on coast, and while the actual wave energy flux reached at least 130kW/m, the simulated one is not more than 60kW/m- significant underestimation.

Fig.3 shows the sea level for Irakli simulated by the storm surge model for Irakli compared with the measurements. Again we observe an underestimation of about 40%.

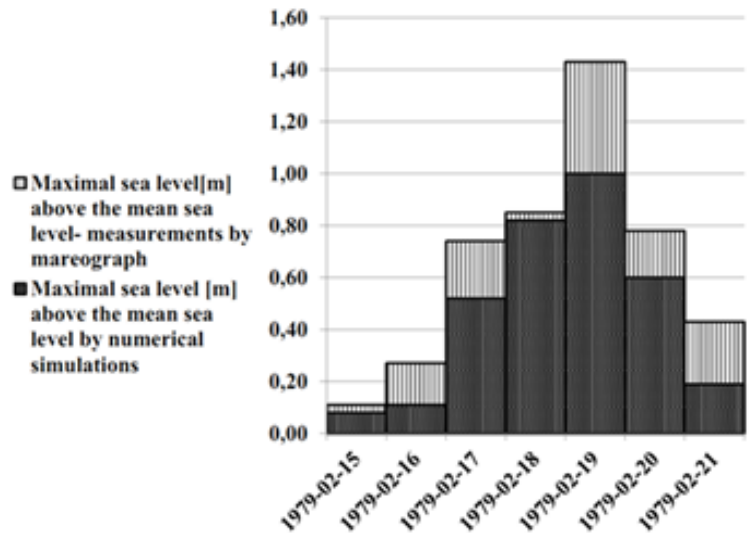

**Fig.3.** Simulation of the sea level near the village of Irakli, compared with the measurement from the tide gauge station there.

## THE CASE 2: FEBRUARY 2003

The second test is for the Eastern part of the Black Sea. The reason for this is the availability at this are of the measurements of the wave parameters and the possibility to compare the wave simulations by SWAN model using two wind inputs: from ERA Interim and from ALADIN model with in-situ measurements in deep water. The case study is for a storm of 01.02.2003. The storm was short, but the significant wave height reached the extreme value of nearly 7m. Fig.4 shows the significant wave height using two different wind input sources. The underestimation when using ERA data is severe (more than 100%).The underestimation using ALADIN model data is about 30% and possibly due to the coarse temporal resolution of the wind input (every 6h at that year- now it is every 3h) and it is also significant but more realistic. The wave energy flux according to the measurements reached 240kW/m, while the simulated with ALADIN input is about 100kW/m and using ERA winds- just 30kW/m. The usage of NCEP reanalysis II leads to overestimation by more than 30%.

## TEST CASE 3: THE STORM OF 06-08 FEBRUARY 2012

The last test case presents a more recent storm- the storm of 06-08 February 2012. The performance of the operational implementation of SWAN wave model for that storm is evaluated, using satellite data [3]. The satellite altimetry data and the visual observations by the coastal meteorological stations of NIMH- BAS show that the significant wave height reached 5m at the southern Bulgarian coast [11]. Fig.5 shows the output from the SWAN model, using wind input ERA data and ALADIN data (0.125° spatial and 3 h temporal resolutions of the input wind fields) near the town of Ahtopol, where the storm caused significant damages.

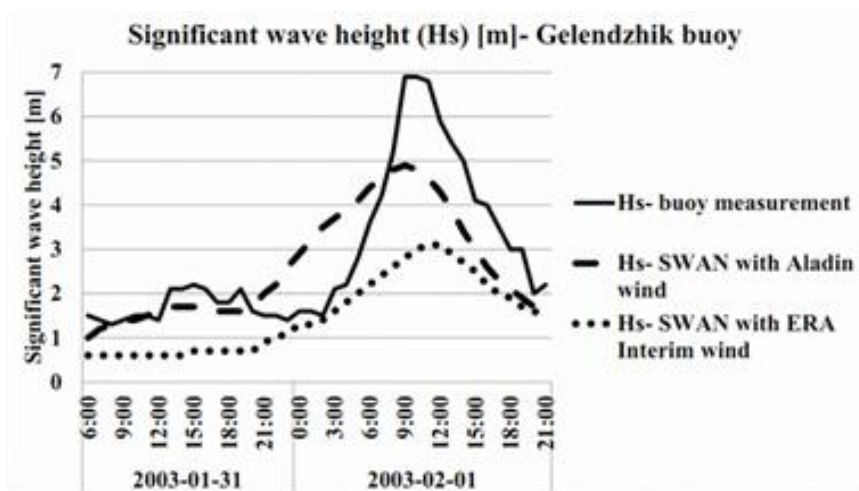

**Fig.4.** SWAN model simulation of the significant wave height, using two different wind inputs, compared with buoy measurements nearby Gelendzhik (Russia)

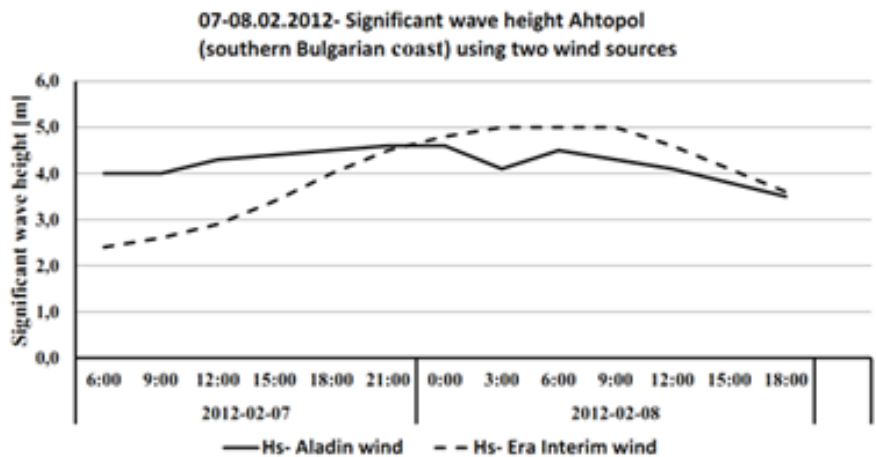

**Fig.5.** Significant wave height- nearshore location nearby the town of Ahtopol, simulated by SWAN model, using ERA and ALADIN wind input for the storm of 06-08.02.2012.

In our previous study [11] we found that the negative bias of the simulated significant wave height by SWAN, using ALADIN wind input for the storms during the winter of 2012 was about 0.3m, therefore the simulated maximum value of 4.7m corresponds to the observations, when it is corrected with the estimated bias. While in the previous two test cases the simulation using ERA Interim wind input showed an underestimation of at least 50%, in this recent case the simulation using ERA input provides a maximum value for the significant wave height comparable with the simulation using ALADIN input. The storm surge simulation for Ahtopol using ALADIN model input data leads to a peak value of +1.3m for the peak of the surge, while it is significantly lower with Era Interim or the operational analysis of ECMWF. However a direct comparison with a tide gauge is impossible, because the tide gauge was destroyed by the storm.

## SUMMARY AND CONCLUSIONS

Three case studies of severe storms in the Black Sea have been presented. For the storm of February 1979 the storm surge and wave heights have been simulated along the Bulgarian coast by the wave and storm surge models, using ERA Interim wind input. The modeled results have been compared with the available measurements. Another test has been pefrormed for a storm of 01.02.2003 in the Eastern Black Sea. Significant wave height has been simulated using SWAN wave model, forced by ERA Interim and ALADIN wind input data. The third test case of 06-08.02.2012 has been done for a recent significant storm at the Western Black Sea coast. For the first two cases it has been found, that the use of ERA Interim coarse resolution wind data leads to a significant underestimation of the waves and storm surge, that can be a major issue.

The overal conclusion is that it is confirmed, that the accuracy of the atmospheric data input for the wave and storm surge models is crucial for their usage for assessment of the natural coastal hazards.In the frame of IncREO project a dynamical downscaling with high resolution of the ERA reanalisis data will be used for a number of selected cases, that can further improve the results especialy for the storms in more distant past.

## ACKNOWLEDGEMENTS

This study has been supported by and performed in the frame of FP7 EU Project IncREO.

## REFERENCES

[1] Dee, D. P., Uppala, S. M., Simmons, A. J., Berrisford, P., Poli, P., Kobayashi, S., Andrae, U., Balmaseda, M. A., Balsamo, G., Bauer, P., Bechtold, P., Beljaars, A. C. M., van de Berg, L., Bidlot, J., Bormann, N., Delsol, C., Dragani, R., Fuentes, M., Geer, A. J., Haimberger, L., Healy, S. B., Hersbach, H., Hólm, E. V., Isaksen, L., Kållberg, P., Köhler, M., Matricardi, M., McNally, A. P., Monge-Sanz, B. M., Morcrette, J.-J., Park, B.-K., Peubey, C., de Rosnay, P., Tavolato, C., Thépaut, J.-N. and Vitart, F. (2011), The ERA-Interim reanalysis: configuration and performance of the data assimilation system. Q.J.R. Meteorol. Soc., 137: 553–597. Doi: 10.1002/qj.828

[2] Booij, N., Ris, R. C., and Holthuijsen, L. H. (1999).A third generation wave model for coastal regions, Part 1: Model description and validation, Journal of Geophysical Research, Vol. 104, No. C4, pp. 7649-7666

[3] Rusu, E. (2010). Modelling of wave–current interactions at the mouths of the Danube. Journal of marine science and technology, 15(2), 143-159

[4] Rusu, E. (2009).Wave energy assessments in the Black Sea. Journal of marine science and technology, 14(3), 359-372.

[5] Rusu, E. (2011). Strategies in using numerical wave models in ocean/coastal applications. Journal of Marine Science and Technology, 19(1), 58-75.

[6] Akpınar, A., van Vledder, G. P., İhsan Kömürcü, M., & Özger, M. (2012). Evaluation of the numerical wave model (SWAN) for wave simulation in the Black Sea. Continental Shelf Research.

[7] Akpınar, A., & Kömürcü, M. İ. (2012). Wave energy potential along the south-east coasts of the Black Sea. Energy,42(1), 289- 302 pp

[8] Rusu, L. (2010). Application of numerical models to evaluate oil spills propagation in the coastal environment of the Black Sea. Journal of Environmental Engineering and Landscape Management, 18(4), 288-295.

[9] Rusu, L. C., & de León Álvarez, S. P. On the Performances of the Third Generation Spectral Wave Models in the Black Sea. ANALELE UNIVERSITĂŢII "DUNĂREA DE JOS" DIN GALAŢI, 23.

[10] Fomin, V. V., Alekseev, D. V., & Ivancha, E. V. E. (2012). Modeling of Wind Waves in the Bays of South-West Part of the Crimea Peninsula. Turkish Journal of Fisheries and Aquatic Sciences, 12, 363-369.

[11] Galabov Vasko, Anna Kortcheva, Marieta Dimitrova, (2012) Towards a system for sea state forecasts in the Bulgarian Black Sea coastal zone: the case of the storm of 07-08 February, 2012,SGEM2012 Conference Proceedings/ ISSN 1314-2704, Vol. 3, 1017 - 1024 pp, DOI: 10.5593/SGEM2012/S14.V3012

[12] Galabov Vasko (2013), On the parameterization of whitecapping and wind input in deep and shallow waters and the strategies for nearshore wave modeling in closed seas, Bulgarian Journal of Meteorology and Hydrology, 18(1-2), 18-37 pp

[13] Mungov, G., & Daniel, P. (2000). Storm surges in the Western Black Sea-operational forecasting. Mediterranean Marine Science, 1(1), 45-50.

[14] Daniel, P. (1997). Forecasting tropical cyclones storm surges at Meteo-France. Computer Modelling of Seas and Coastal Regions III, 119-128.

[15] Bogatchev A (2008), Changes in operational suite of ALADIN –BG, ALADIN Newsletter No 34, available online at http://www.cnrm.meteo.fr/aladin/spip.php?article111

[16] Andreeva N., N. Valchev, E.Trifonova, P.Eftimova, D.Kirilova, M.Georgieva (2011), Literary review of historical storm events in the western Black Sea, Proceedings of the Union of scientists Varna, 105-112pp, available online at http://su-varna.org/izdanij/MNauki-2011/pages_105_112.pdf, (In Bulgarian)

[17] Belberov, Z., E. Trifonova, N. Valchev, N. Andreeva, P. Eftimova, (2009), Contemporary reconstruction of the historical storm of February 1979 and assessment of its impact on the coastal zone infrastructure, Proc. of the "International Multidisciplinary Scientific Geo-Conference – SGEM'2009", vol. 2, pp. 243-250